\documentclass{llncs}
\usepackage{graphicx}
\usepackage{multirow}
\usepackage{mathtools}
\usepackage{diagbox}
\usepackage{amssymb}
\begin{document}
%

\title{Network Sampling Using K-hop Random Walks for Heterogeneous Network Embedding}


%
%
\author{Akash Anil\inst{1}\and
Ajay Ladhar\inst{2}\and
Sandeep Singh\inst{2}\and
Uppinder Chugh\inst{1}\and
Sanasam Ranbir Singh\inst{1}}

\institute{Indian Institute of Technology Guwahati, Assam, India \and
National Institute of Technology Silchar, Assam, India\\ 
\email{<a.anil.iitg, ajayladhar7, singhsandeep138, uppinderchugh>@gmail.com, ranbir@iitg.ac.in}}

%
%


%
\maketitle              
\begin{abstract}
Sampling a network is an important prerequisite for unsupervised network embedding. 
Further, random walk has widely been used for sampling in previous studies.
Since random walk based sampling tends to traverse adjacent neighbors, it may not be suitable for heterogeneous network because in heterogeneous networks two adjacent nodes often belong to different types. Therefore, this paper proposes a K-hop random walk based sampling approach which includes a node in the sample list only if it is separated by $K$ hops from the source node.
We exploit the samples generated using K-hop random walker for network embedding using skip-gram model ({\tt word2vec}). Thereafter, the performance of network embedding is evaluated on co-authorship prediction task in heterogeneous DBLP network. We compare the efficacy of network embedding exploiting proposed sampling approach with recently proposed best performing network embedding models namely, {\tt Metapath2vec} and {\tt Node2vec}. It is evident that the proposed sampling approach yields better quality of embeddings and out-performs baselines in majority of the cases.

\end{abstract}

\section{Introduction}
\label{sec:introduction}
Network embedding using Artificial Neural network over social and information networks became popular in a very recent time. 
Majority of the unsupervised network embedding methods use a two-step framework, \textit{(i) generate the network samples, and (ii) train a neural network on these samples to generate embeddings}~\cite{tang2015line,Grover:2016:NSF:2939672.2939754,dong2017metapath2vec,perozzi2014deepwalk}. 

Random walk is one of the popular and scalable sampling methods for real-world information networks. To generate network samples, early network embedding models use the first-order random walk where the transition to next node only depends on the current node~\cite{perozzi2014deepwalk,tang2015line}. However, in real-world information networks, a second-order random walk (transition to next node depends on current and previous nodes) preserves
network characteristics more efficiently by capturing community dynamics and structural equivalences~\cite{Grover:2016:NSF:2939672.2939754,wu2018second}. 

The above-discussed random walk based sampling methods may be efficient for homogeneous network (consists singular type of nodes and edges) but may fail for heterogeneous network (consists multiple types of nodes and edges). It is because sampling methods exploiting random walk (first or second order) are bound to consider adjacent neighbors. However, in general, heterogeneous network is often constructed such that two nodes having similar characteristics are connected via the adjacent neighbors. For example, in a heterogeneous bibliographic network consisting of \textit{author, paper, and venue} as nodes, two \textit{authors} may be connected via a \textit{paper} or \textit{venue}. Therefore, incorporating adjacent neighbors in sampling a heterogeneous network may degrade the network embedding quality. On the other hand, distant neighbors in heterogeneous network may preserve the mutual proximity better and improve the embedding quality.   
Motivated with this, in this paper, we propose a K-hop random walk based sampling approach (RW-$K$) which randomly samples two nodes together only if there exist $K$ nodes or hops between them.

We use the RW-$K$ based network samples for network embedding using single layer skip-gram based neural network model, i.e. {\tt word2vec}~\cite{mikolov2013distributed}. Further, we evaluate the embedding performance over co-authorship prediction task in heterogeneous DBLP bibliographic\footnote{https://dblp.uni-trier.de/} network. We compare the performance of 
embedding using the proposed RW-$K$ sampling method to recent embedding methods exploiting random walk based sampling namely \texttt{Metapath2vec}~\cite{dong2017metapath2vec} and \texttt{Node2vec}~\cite{Grover:2016:NSF:2939672.2939754}.  From various experiments, it is evident that by setting hop $K=1$ and $K=2$, RW-$K$ helps in improving the embedding quality and out-performs the baselines in 75\% times.


\section{Network Sampling using K-hop Random Walk}
\label{sec:sampling}
From the perspective of computer network, a hop is defined as maximum number of intermediate devices through which data must pass from source to destination. Let, $G(V,E)$ is a graph representing the underlying network, $K \in \mathbb{Z}$ is the hop size, $u \in V$  be the source, $v \in V$ be the destination, then RW-$K$ samples $(u,v)$ if there exist $K$ number of nodes between the path $(u,v)$.

\section{Experimental Analysis}
\label{sec:experimental_analysis}

\subsection{Dataset}
\label{sec:dataset}
This paper uses the DBLP bibliographic information from 1968 to 2011 processed in the study~\cite{yang2012topic}. From this data, we consider Author, Paper, and Venue as node types and corresponding relationships among these node types to generate a heterogeneous bibliographic network. Further, this network is split into two parts i.e. (i) from 1968 to 2008 for training the network embedding model and (ii) from 2009 to 2011 for evaluating the performance of embeddings on co-authorship prediction task.  


\subsection{Experimental Result and Discussion}
\label{sec:experimental_setups}
To generate the network samples, we investigate with the different values of $K$ in RW-$K$ i.e. 0, 1, 2,..., etc. We observe that $K>2$ does not improve the embedding quality significantly. Thus, we consider $K$ only upto 2 and generate the samples (maximum of $100$ in length) by iterating 30 times for any given node in the network\footnote{We investigated with higher iteration value i.e. 40, 50, etc. and observed that higher than 30 iterations do not show much difference.}. Now, for network embedding, we exploit the skip-gram with negative sampling neural network framework ({\tt word2vec}~\cite{mikolov2013distributed}) on these samples. 

We evaluate the quality of network embedding on predicting co-authorship task. Therefore, we consider all the new co-authorship relations appeared in 2009-2011 for the authors present in training network as test links.  
Similar to study~\cite{Tsitsulin:2018:VVG:3178876.3186120}, we map the co-authorship prediction task from link prediction to a binary classification task.
Further, we generate an equal number of negative test links on the nodes appeared in the above test links. Similar to studies~\cite{Grover:2016:NSF:2939672.2939754,Tsitsulin:2018:VVG:3178876.3186120}, we generate the edge feature using Hadamard operator on the node embeddings. 
 We use four state-of-art classifiers namely, Decision Tree (DT), Naive Bayes (NB), Random Forest (RF), and Logistic Regression (LR) on a random 80:20 train-test split over combined test links (actual test links with randomly generated negative test links). Thereafter, we repeat the same setup for 10 times and the average AUC is considered to assess the performance of co-authorship prediction. Further, we compare the performance of co-authorship prediction by network embedding exploiting RW-$K$ with recently proposed network embedding models namely \texttt{Node2vec} and \texttt{Metapath2vec} exploiting random walk for sampling the network.

\begin{table*}[t]
\caption{AUC score for Co-Authorship Prediction Task by different Classifiers}
\label{tab:result}
\resizebox{\textwidth}{!}{
\begin{tabular}{|c|c|c|c|c|c|c|}
\hline
\textbf{Classifier} & \textbf{Metapath2vec} & \textbf{Node2vec} & \textbf{RW-(K=0)} & \textbf{RW-(K=1)} & \textbf{RW-(K=2)} & \textbf{Concat(K=0, K=1, K=2)} \\ \hline
\textbf{DT} & 0.644 & 0.657 & 0.647 & 0.673 & 0.673 & 0.679 \\ \hline
\textbf{NB} & 0.717 & 0.719 & 0.690 & 0.718 & 0.713 & 0.718 \\ \hline
\textbf{RF} & 0.72 & 0.73 & 0.721 & 0.745 & 0.750 & 0.752 \\ \hline
\textbf{LR} & 0.766 & 0.777 & 0.766 & 0.780 & 0.792 & \textbf{0.804} \\ \hline
\end{tabular}
}
\end{table*}

From Table~\ref{tab:result}, it is evident that network embedding exploiting RW-$K$ outperforms the state-of-art baselines for atleast three out of four classifiers. However, we observe that proposed model exploiting adjacent neighbors ($K=0$) performs poorly and comparable to {\tt Metapath2vec}. Further, it is clearly visible that the proposed model with $K=1$ performs better than \texttt{Metapath2vec, and Node2vec} for three out of four classifiers. This observation is also consistent with $K=2$. Thus, it can be inferred that network sampling in a heterogeneous network may be better by incorporating random nodes together if they are separated by some hops. 

Among the proposed models, RW with $K=2$ yields better performance for almost all the classifiers. Thus, it can be inferred that in heterogeneous bibliographic network, distant neighbors may preserve the network characteristics better than the adjacent neighbors. 

Further, to assess the combined contributions by different hops, we again evaluate the co-authorship prediction performance by concatenating node embeddings for $K=0, K=1$, and $K=2$. We observe that concatenation of embeddings from different hops improves the co-authorship prediction performance. 
Hence, it can be inferred that different hop sizes for RW-$K$ preserves rich network characteristics and should be selected carefully.

\section{Conclusion}
\label{sec:conclusion}
This paper studies the problem of network sampling for heterogeneous network embedding. We propose a K-hop random walk based sampling approach which generates more meaningful network samples. Further, these network samples are used to generate network embedding using the {\tt word2vec} framework.
The efficacy of proposed sampling approach in network embedding is evaluated on co-authorship prediction task in heterogeneous DBLP bibliographic network and compared to suitable baselines. We observe that the proposed sampling approach out-performs the baselines in majority of the cases. 

The limitation of this work is that it requires a hyper-parameter $K$ as hop size. To alleviate this limitation, learning the optimal hop size from supervised knowledge of the underlying network may be a future direction.    

\bibliographystyle{splncs04} 
\bibliography{ref_ecir}

\end{document}